\documentclass[a4paper,english,aps,manuscript]{revtex4}
\usepackage[T1]{fontenc}
\usepackage[latin1]{inputenc}
\usepackage{amssymb}

\makeatletter

\newcommand{\noun}[1]{\textsc{#1}}

\usepackage{babel}
\makeatother
\begin{document}

\title{Delayed response of a Fermion Pair Condensate to a modulation of
the interaction strength}

\author{J. Plata}

\affiliation{Departamento de F\'{\i}sica Fundamental II, Universidad de La Laguna,\\
 La Laguna E38204, Tenerife, Spain.}

\begin{abstract}
The effect of a sinusoidal modulation of the interaction strength
on a Fermion pair condensate is analytically studied. The system is
described by a generalization of the coupled fermion-boson model that
incorporates a time-dependent intermode coupling induced via a magnetic
Feshbach resonance. Nontrivial effects are shown to emerge depending
on the relative magnitude of the modulation period and the relaxation
time of the condensate. Specifically, a nonadiabatic modulation drives
the system out of thermal equilibrium: the external field induces
a variation of the quasiparticle energies, and, in turn, a disequilibrium
of the associated populations. The subsequent relaxation process is
studied and an analytical description of the gap dynamics is obtained.
Recent experimental findings are explained: the delay observed in
the response to the applied field is understood as a temperature effect
linked to the condensate relaxation time. 
\end{abstract}

\pacs{03.75Ss, 05.30.Fk}

\maketitle
The study of ultracold atomic gases has led to a remarkable series
of experimental realizations of fundamental effects \cite{key-1}.
Essential to many of these achievements has been the control of the
interaction strength via a Feshbach resonance (FR), which has allowed
the emergence of these systems as a practical testing ground for quantum-statistical
and many-body physics. Specially relevant has been the realization,
with a two-component Fermi gas of atoms, of the crossover from a molecular
Bose-Einstein condensate (BEC) to a Bardeen-Cooper-Schrieffer (BCS)
superfluid of loosely bound atom pairs \cite{key-2,key-3,key-4,key-5,key-6,key-7,key-8,key-9,key-10,key-11}.
By applying a magnetic FR, the interaction is led to change from repulsive
in the BEC phase to attractive in the BCS side. Whereas two-body physics
supports a bound molecular state in the BEC side, the formation of
pairs in the BCS regime occurs only due to many-body effects. Significant
advances have been made in the understanding of different aspects
of this transition. In particular, the role of thermal fluctuations
has been extensively analyzed \cite{key-12}. Despite those advances,
further work on the characterization of nonequilibrium aspects of
the crossover is required. Indeed\emph{,} the extension of the experiments
on scattering-length variations to unexplored time-dependent regimes
and the setup of an expanded theoretical framework where the emergent
effects can be understood are current challenges. Here, we aim at
explaining recent experimental results on the nontrivial dynamics
resulting from a sinusoidal modulation of the interaction strength
\cite{key-13}. This will require a precise characterization of the
different time scales, in particular, of the relaxation time of the
condensate. The interest of the study is not restricted to the field
of ultracold atomic gases. In fact, the expansion of the BCS model
proposed in our approach can have broad applicability: the effect
of a disequilibrium population on the gap dynamics, which is the central
issue in the analysis, is relevant to topics ranging from nonequilibrium
superconductivity \cite{key-14,key-15} to quench dynamics in superfluid
$^{3}He$ \cite{key-16}.

As starting point we take the experiments reported in Ref. {[}13{]}.
In them, the interaction strength of a gas of ultracold $^{6}Li$
atoms in the BCS regime was sinusoidally modulated through a magnetic
FR. (The (broad) FR at 834 G between the two lowest hyperfine states
was used.) The system response, which consisted in a damped oscillation
of the condensate fraction with the modulation frequency, was found
to be delayed with respect to the applied field. The delay showed
no appreciable changes at different cycles of the external field;
moreover, it presented the same scale for widely different frequencies.
Additionally, the damping time was observed to be much longer than
the driving period. In a preliminary analysis, the deferred response
was conjectured to be rooted in the finite relaxation time of the
condensate; furthermore, the decay of the oscillation amplitude was
linked to heating resulting from the nonadiabaticity (on the gap time
scale) of the process. (In the present paper, {}``adiabatic'' will
also be applied to a modulation much slower than the condensate relaxation.)
To evaluate those conjectures, the measured characteristic times were
compared with related theoretical predictions. However, that analysis
was not conclusive about  the origin of the recorded behavior because
of the limitations of the available models, specially, of the lack
of an appropriate description of finite-temperature effects. Our objective
is to provide a theoretical framework where the observed features
can be understood, and, in particular, the previous conjectures can
be assessed. To this end, we concentrate on conditions which allow
an analytical description of the dynamics, and, consequently, a clear
identification of the dominant mechanisms.  

We consider a gas of ultracold Fermi atoms with two hyperfine states
coupled to a molecular two-particle state via a magnetic FR. Our methodology
to deal with magnetic-field modulations combines three main elements.
First, the standard framework, namely, the coupled fermion-boson model
\cite{key-17,key-18,key-19,key-20}, is expanded by incorporating
a time-dependent intermode coupling. Second, a perturbative scheme,
valid for a sufficiently small modulation amplitude, is introduced
in the Hartree-Fock-Bogoliubov (HFB) description. In this approach,
finite-temperature effects are tackled and the out-of-equilibrium
situation induced by the modulation is characterized. Finally, a method
for describing the evolution of the quasiparticle populations, and,
subsequently, the gap dynamics is presented. Accordingly, we start
from the unmodulated system: the grand-canonical Hamiltonian reads 

\begin{eqnarray}
H-\mu N & = & \sum_{\mathbf{k},\sigma}\varepsilon_{\mathbf{k}}a_{\mathbf{k},\sigma}^{\dagger}a_{\mathbf{k},\sigma}+V_{int}\sum_{\mathbf{q,k,k^{\prime}}}a_{\frac{\mathbf{q}}{2}+\mathbf{k},\uparrow}^{\dagger}a_{\frac{\mathbf{q}}{2}-\mathbf{k},\downarrow}^{\dagger}a_{\frac{\mathbf{q}}{2}-\mathbf{k^{\prime}},\downarrow}a_{\frac{\mathbf{q}}{2}+\mathbf{k^{\prime}},\uparrow}+\nonumber \\
 &  & \sum_{\mathbf{q}}\left(\varepsilon_{\mathbf{q}}^{m}+\hbar\nu_{0}\right)b_{\mathbf{q}}^{\dagger}b_{\mathbf{q}}+g\sum_{\mathbf{q,k}}\left(b_{\mathbf{q}}a_{\frac{\mathbf{q}}{2}+\mathbf{k},\uparrow}^{\dagger}a_{\frac{\mathbf{q}}{2}-\mathbf{k},\downarrow}^{\dagger}+\textnormal{h.c.}\right)\end{eqnarray}
where $\mu$ is the chemical potential, $N$ is the total number of
bare Fermi atoms, $a_{\mathbf{k},\sigma}^{\dagger}$ ($a_{\mathbf{k},\sigma}$)
denotes a fermionic creation (annihilation) operator of an atom with
momentum $\mathbf{k}$ and spin $\sigma$, ($\sigma\in\left\{ \uparrow,\downarrow\right\} $),
and $b_{\mathbf{q}}^{\dagger}$ ($b_{\mathbf{q}}$) is a bosonic operator
that creates (destroys) a molecule with momentum $\mathbf{q}$. The
populations corresponding to the two hyperfine states are assumed
to be equal. The free dispersion relations for fermions and bosons
are $\varepsilon_{\mathbf{k}}=\hbar^{2}\mathbf{k}^{2}/2m-\mu$ and
$\varepsilon_{\mathbf{q}}^{m}=\hbar^{2}q^{2}/4m-2\mu$, respectively.
The binary attractive interaction potential between fermions is characterized
by $V_{int}$(<$0$). Additionally, $g$ represents the FR coupling
between the closed and the open channel states, $\nu_{0}$ being the
detuning of the boson resonance state from the collision continuum.

Initially, the system is at equilibrium at a finite temperature $T$.
In that situation, a sinusoidal modulation of the detuning from the
FR is applied. Correspondingly, $\nu_{0}$ is replaced by $\nu(t)=\nu_{0}+A\sin\omega_{p}t$.
It is assumed that $V_{int}$, which characterizes the attractive
pairing interaction resulting from nonresonant processes, is not affected
by the applied detuning of the FR\emph{.} Through the unitary transformation
$U(t)=e^{i\frac{A}{\omega_{p}}\cos\omega_{p}t\sum_{\mathbf{q}}b_{\mathbf{q}}^{\dagger}b_{\mathbf{q}}}$,
the Hamiltonian is transformed into $H^{\prime}=U^{\dagger}HU-i\hbar U^{\dagger}\dot{U}$;
consequently, Eq. (1), {[}with $\nu_{0}$ replaced by $\nu(t)${]},
is rewritten as 

\begin{eqnarray}
H^{\prime}-\mu N & = & \sum_{\mathbf{k},\sigma}\varepsilon_{\mathbf{k}}a_{\mathbf{k},\sigma}^{\dagger}a_{\mathbf{k},\sigma}+V_{int}\sum_{\mathbf{q,k,k^{\prime}}}a_{\frac{\mathbf{q}}{2}+\mathbf{k},\uparrow}^{\dagger}a_{\frac{\mathbf{q}}{2}-\mathbf{k},\downarrow}^{\dagger}a_{\frac{\mathbf{q}}{2}-\mathbf{k^{\prime}},\downarrow}a_{\frac{\mathbf{q}}{2}+\mathbf{k^{\prime}},\uparrow}+\nonumber \\
 &  & \sum_{\mathbf{q}}(\varepsilon_{\mathbf{q}}^{m}+\hbar\nu_{0})b_{\mathbf{q}}^{\dagger}b_{\mathbf{q}}+\left(g\sum_{\mathbf{q,k}}e^{i\frac{A}{\omega_{p}}\cos\omega_{p}t}b_{\mathbf{q}}a_{\frac{\mathbf{q}}{2}+\mathbf{k},\uparrow}^{\dagger}a_{\frac{\mathbf{q}}{2}-\mathbf{k},\downarrow}^{\dagger}+\textnormal{h.c.}\right).\end{eqnarray}
 Our procedure to analyze the dynamics starts, like the standard HFB
approach \cite{key-20,key-18}, with the introduction of three mean
fields: $n\equiv\sum_{\mathbf{k}}\left\langle a_{\mathbf{k},\sigma}^{\dagger}a_{\mathbf{k},\sigma}\right\rangle $
for the spin density, $\Delta\equiv\left|V_{int}\right|\sum_{\mathbf{k}}\left\langle a_{-\mathbf{k},\downarrow}a_{\mathbf{k},\uparrow}\right\rangle $
for the pairing field, and $\phi_{m}\equiv\left\langle b_{\mathbf{q=0}}\right\rangle $
for the boson field. (We take $\mathbf{q}=\mathbf{0}$ as we focus
on the condensed molecular field.) Next, a perturbative scheme is
set up as follows. From Eq. (2), it is apparent that the effect of
the magnetic modulation can be understood as a time variation of the
intermode coupling strength: we can work with the effective strength
$g_{eff}\equiv ge^{i\frac{A}{\omega_{p}}\cos\omega_{p}t}=g+\delta g(t)$,
where $\delta g(t)$ stands for the modulation-induced increment.
(Note that the time dependence of $g_{eff}$ prevents the effective
one-channel reduction applicable, for a broad FR, to the undriven
coupled fermion-boson model.) Furthermore, from the expansion $e^{i\frac{A}{\omega_{p}}\cos\omega_{p}t}=\sum_{l=-\infty}^{\infty}i^{l}J_{l}(A/\omega_{p})e^{il\omega_{p}t}$,
and taking into account the properties of the Bessel functions, it
follows that, for $A/\omega_{p}\ll1$, we can make the approximation
$\delta g\simeq i2gJ_{1}(A/\omega_{p})\cos\omega_{p}t$, the magnitude
of $\delta g$ being much smaller than that of $g$. (Higher-order
terms will be discussed later on.) In turn, the previously defined
mean fields can be expressed as $n=n_{0}+\delta n$, $\Delta=\Delta_{0}+\delta\Delta$,
and $\phi_{m}=\phi_{m,0}+\delta\phi_{m}$, where $n_{0}$, $\Delta_{0}$,
and $\phi_{m,0}$ are the respective values in the absence of the
magnetic variation, and, $\delta n$, $\delta\Delta$, and $\delta\phi_{m}$
are the corresponding modulation-induced increments. To first order,
the complete Hamiltonian can be split as $H^{\prime}-\mu N\simeq H_{0}+H_{per}$.
The zero-order term, which describes the unmodulated system, is given
by $H_{0}=\sum_{\mathbf{k},\sigma}V_{k}a_{\mathbf{k},\sigma}^{\dagger}a_{\mathbf{k},\sigma}-\sum_{\mathbf{k}}(\tilde{\Delta}_{0}a_{\mathbf{k},\uparrow}^{\dagger}a_{-\mathbf{k},\downarrow}^{\dagger}+\textnormal{h.c.})$
and corresponds to an effective BCS model with mode energy $V_{k}\equiv\varepsilon_{\mathbf{k}}-\mu+V_{int}n_{0}$
and gap $\tilde{\Delta}_{0}\equiv\Delta_{0}-g\phi_{m,0}$. The first-order
correction reads $H_{per}=-\delta\tilde{\Delta}(t)\sum_{\mathbf{k}}a_{\mathbf{k},\uparrow}^{\dagger}a_{-\mathbf{k},\downarrow}^{\dagger}+\textnormal{h.c.},$
where $\delta\tilde{\Delta}(t)=\delta\Delta-g\delta\phi_{m}-\delta g\phi_{m,0}$
is the increment of the generalized order parameter $\tilde{\Delta}\equiv\Delta-g\phi_{m}$.
(We have neglected the variation in the atomic density, i.e., we have
taken $\delta n\simeq0$, which is justified for a broad FR \cite{key-21}.
The generalization required to deal with a narrow FR is straightforward.)
To complete the mean-field description, we must add the equation for
the evolution of the boson mode, namely, \textbf{\noun{}}\textbf{$i\hbar\frac{d(\phi_{m,0}+\delta\phi_{m})}{dt}=(\nu_{0}-2\mu)(\phi_{m,0}+\delta\phi_{m})+\frac{g_{eff}}{\left|V_{int}\right|}(\Delta_{0}+\delta\Delta)$},
which, through the application of our perturbative approach and taking
into account that the equilibrium molecular field is given by $\phi_{m,0}=\frac{g\Delta_{0}}{\left|V_{int}\right|(2\mu-\nu_{0})}$,
is converted into 

\begin{eqnarray}
i\hbar\frac{d\delta\phi_{m}}{dt} & = & (\nu_{0}-2\mu)\delta\phi_{m}+\frac{g}{\left|V_{int}\right|}\delta\Delta+\frac{\delta g}{\left|V_{int}\right|}\Delta_{0}.\end{eqnarray}
Here, the presence of the driving term $\frac{\delta g}{\left|V_{int}\right|}\Delta_{0}$
points to the oscillation of $\delta\phi_{m}$ with frequency $\omega_{p}$.

$H_{0}$ is standardly diagonalized through a Bogoliubov transformation
(BT) characterized by the fermionic operators $c_{\mathbf{k},\uparrow}=\cos\theta_{k}a_{\mathbf{k},\uparrow}-\sin\theta_{k}a_{-\mathbf{k},\downarrow}^{\dagger}$
and $c_{-\mathbf{\mathbf{k},\downarrow}}^{\dagger}=\sin\theta_{k}a_{\mathbf{k},\uparrow}+\cos\theta_{k}a_{-\mathbf{k},\downarrow}^{\dagger}$,
where $\theta_{k}$ is defined by $\tan(2\theta_{k})=\left|\tilde{\Delta}_{0}\right|/V_{k}$
\cite{key-18,key-20}. With this BT, $H_{0}$ is cast into $H_{0}=\sum_{\mathbf{k}}E_{k,0}(c_{\mathbf{k},\uparrow}^{\dagger}c_{\mathbf{k},\uparrow}+c_{\mathbf{\mathbf{k},\downarrow}}^{\dagger}c_{\mathbf{\mathbf{k},\downarrow}})+\textrm{constant}.$
Importantly, $c_{\mathbf{k},\uparrow}^{\dagger}$ ($c_{\mathbf{k},\uparrow}$)
corresponds to the creation (annihilation) operator of a quasi-particle
excitation with momentum $\mathbf{k}$ and spin $\uparrow$ from the
BCS state, which acts as an effective vacuum state. The associated
excitation energies are $E_{k,0}=\sqrt{V_{k}^{2}+\tilde{\Delta}_{0}^{2}}$.
The excitation gap $\tilde{\Delta}_{0}$ combines the mean field $\Delta_{0}$,
obtained from the BCS equation $\Delta_{0}=\frac{\left|V_{int}\right|}{2}\sum_{\mathbf{k}}(2f_{k}-1)\sin2\theta_{k}$,
with the equilibrium molecular field $\phi_{m,0}$, which results
from the equation for the boson mode in the absence of driving. As
thermal equilibrium is assumed for the system before the application
of the magnetic modulation, the populations of the quasiparticle states
$\left\{ f_{k}\right\} $ are initially given by the Fermi distribution
function $f_{k,0}^{eq}=1/(1+e^{E_{k,0}/k_{B}T})$. 

Through the previously defined BT, the perturbation Hamiltonian is
converted into

\begin{eqnarray}
H_{per} & = & \sum_{\mathbf{k}}\delta E_{k}(t)\left(c_{\mathbf{k},\uparrow}^{\dagger}c_{\mathbf{k},\uparrow}+c_{\mathbf{\mathbf{k},\downarrow}}^{\dagger}c_{\mathbf{\mathbf{k},\downarrow}}\right)+\left(G_{k}(t)c_{\mathbf{k},\uparrow}^{\dagger}c_{-\mathbf{k},\downarrow}^{\dagger}+\textnormal{h.c.}\right),\end{eqnarray}
where $\delta E_{k}(t)\equiv\frac{1}{2}\delta\tilde{\Delta}(t)\sin2\theta_{k}+\textnormal{c.c.}$,
and $G_{k}(t)\equiv-\delta\tilde{\Delta}(t)\cos^{2}\theta_{k}+\delta\tilde{\Delta}^{*}(t)\sin^{2}\theta_{k}$.
From the form of $H_{per}$, a first picture of the dynamical implications
of the field modulation can be drawn. The (time-dependent) diagonal
terms lead to a time variation of the quasiparticle energies, which
become $E_{k}(t)=E_{k,0}+\delta E_{k}(t)$. The non-diagonal terms
represent modulation-induced interactions between the vacuum state
and a doubly-excited state. Importantly, these coupling terms, which
oscillate with the external frequency $\omega_{p}$, are relevant
only when they can induce an effective resonance between the BCS state
and the two-excitation configuration, i.e., only when $\omega_{p}\geq2\tilde{\Delta}_{0}$,
($\hbar=1$.) Here, in order to isolate the mechanisms responsible
for the delayed response, we concentrate first on the regime defined
by $\omega_{p}<2\tilde{\Delta}_{0}$. In this frequency range, the
interaction terms can be discarded, and the perturbation Hamiltonian
can be approximated as $H_{per}=\sum_{\mathbf{k}}\delta E_{k}(c_{\mathbf{k},\uparrow}^{\dagger}c_{\mathbf{k},\uparrow}+c_{\mathbf{\mathbf{k},\downarrow}}^{\dagger}c_{\mathbf{\mathbf{k},\downarrow}})$.
Hence, the quasiparticle states of the unmodulated system still provide
a diagonal representation of the driven Hamiltonian. The appearance
of heating effects outside this regime will be discussed later on. 

Now we turn to incorporate finite-temperature effects in the above
framework. The modulation drives the system out of equilibrium as
the initial populations, i.e., the thermal values associated with
the unmodulated energies, do not correspond to the Fermi distribution
$f_{k}^{eq}(t)=1/(1+e^{E_{k}(t)/k_{B}T})$ for the actual (time-varying)
energies. The description of the effect of this quasi-particle disequilibrium
on the gap dynamics requires a selfconsistent approach since the energies
and the gap are interdependent. Indeed, as shown by the expression
$E_{k}(t)=E_{k}^{(0)}+\left(\frac{\tilde{\Delta}_{0}}{2E_{k,0}}\delta\tilde{\Delta}(t)+\textnormal{c.c.}\right)$,
the quasi-particle energies are affected by the gap evolution and
by the molecular-field variation; in turn, the $\left\{ E_{k}(t)\right\} $
enter the general gap equation \cite{key-18,key-20}, 

\begin{eqnarray}
\Delta(t) & =\frac{\left|V_{int}\right|}{2} & \sum_{\mathbf{k}}\left[2f_{k}(t)-1\right]\sin2\theta_{k},\end{eqnarray}
via the (changing) associated populations $\left\{ f_{k}(t)\right\} $.
We will see that it is precisely the evolution of the populations,
more specifically, their relaxation towards equilibrium, that gives
the keys to understanding the experimental results. An important aspect
of this problem can be understood by now: a finite relaxation time
$\tau_{f}$ of the $\left\{ f_{k}\right\} $ is necessary for the
appearance of the gap delay. In fact, for a sudden relaxation, the
populations follow adiabatically (on the relaxation time scale) the
equilibrium values $\left\{ f_{k}^{eq}(t)\right\} $ associated with
the time-dependent energies. The evolution corresponds then to a sequence
of equilibrium states where time enters as a parameter, the associated
gap dynamics being {}``trivial'': no delay between the gap evolution
and the external field emerges. Therefore, to reproduce the delayed
response, we must go beyond that adiabatic regime. Accordingly, we
present a self-contained derivation of the dynamics with no constraints
on time scales. The evolution of the populations is assumed to be
governed by the equation \cite{key-14}

\begin{eqnarray}
\frac{df_{k}}{dt} & = & -\frac{1}{\tau_{f}}\left[f_{k}(t)-f_{k}^{eq}(t)\right],\end{eqnarray}
where $1/\tau_{f}$ represents the effective thermalization rate.
The relaxation mechanism can be conjectured to be rooted in collisions
between excited particles. Here, we do not go into details of the
dependence of $\tau_{f}$ on the system characteristics; instead,
as we focus on general aspects of the role of the population thermalization
in the condensate relaxation, we consider a generic $\tau_{f}$. Here,
a comment on the stability of the temperature is in order. One must
take into account that the standard trapping conditions allow assuming
the stability of the temperature for the considered small variations
of the scattering length. In this sense, we recall that a grand-canonical
description, which is routinely applied in this context as it can
incorporate the possible exchange of particles between the condensate
fraction and the thermal cloud, implies that a fixed temperature can
be reasonably assumed. Also, it is worth stressing that a parallel
treatment of the fluctuations of the condensate field would be necessary
to formally complete our description. However, it is shown that, in
our perturbative regime, the non-condensate fraction has a second-order
effect on the gap dynamics. (For a systematic treatment of different
aspects of the role of fluctuations, see Ref. {[}12{]}.)

Eq. (6) is exactly solved to give $f_{k}(t)=f_{k}^{eq}(t)-\int_{-\infty}^{t}e^{-(t-t^{\prime})/\tau_{f}}\frac{df_{k}^{eq}}{dt}(t^{\prime})dt^{\prime}.$
Next, this expression for the populations is introduced into Eq. (5)
to give the following integral-differential equation for the order
parameter

\emph{\begin{eqnarray}
\Delta_{0}+\delta\Delta(t) & =\frac{\left|V_{int}\right|}{2} & \sum_{\mathbf{k}}\left[2\left(f_{k}^{eq}(t)-\int_{-\infty}^{t}e^{-(t-t^{\prime})/\tau_{f}}\frac{df_{k}^{eq}}{dt}(t^{\prime})dt^{\prime}\right)-1\right]\sin2\theta_{k}.\end{eqnarray}
}Here, one must take into account that $f_{k}^{eq}(t)$ contains $\delta\Delta(t)$
and $\delta\phi_{m}$, and $\frac{df_{k}^{eq}}{dt}$ contains $\dot{\delta\Delta}$
and $\dot{\delta\phi_{m}}$. Hence, we have obtained a description
of the gap evolution, albeit in implicit form. Given the complexity
of this picture, the problem of identifying the origin of the deferred
response, reduced at this point of the study to that of uncovering
the connection between the delay time and $\tau_{f}$, is still nontrivial.
However, it simplifies considerably in the following regime, where
an explicit characterization of the gap evolution is feasible. Specifically,
for $E_{k}\sim\Delta\ll T\approx T_{c}$, ($k_{B}=1$), where $T_{c}$
is the temperature for the BCS transition, we can make the approximations
$f_{k}^{eq}(t)\simeq f_{k,0}^{eq}+\frac{df_{k}^{eq}}{dE_{k}}\delta E_{k}(t)$
and $\frac{df_{k}^{eq}}{dE_{k}}\simeq-\frac{1}{4T_{c}}$. Then, with
the expression for the unperturbed gap $\Delta_{0}=\frac{\left|V_{int}\right|}{2}\sum_{\mathbf{k}}\left[2f_{k,0}^{eq}-1\right]\sin2\theta_{k}$
and the approximation $\frac{df_{k}^{eq}}{dt}\simeq-\frac{1}{4T_{c}}\frac{dE_{k}}{dt}=-\frac{1}{4T_{c}}\sin2\theta_{k}(\dot{\delta\Delta}-g\dot{\delta\phi_{m}})$,
Eq. (7) is cast into\emph{\begin{eqnarray}
\delta\Delta & =-\frac{\left|V_{int}\right|}{8T_{c}} & \sum_{\mathbf{k}}\sin^{2}2\theta_{k}\left[(\delta\Delta-g\delta\phi_{m})-\int_{-\infty}^{t}e^{-(t-t^{\prime})/\tau_{f}}(\dot{\delta\Delta}-g\dot{\delta\phi_{m}})dt^{\prime}\right].\end{eqnarray}
}Now, following a standard procedure, this integral-differential equation
is converted into the differential equation \begin{equation}
\frac{d\delta\Delta}{dt}=-\frac{1-\chi}{\tau_{f}}\delta\Delta-\frac{\chi}{\tau_{f}}g\delta\phi_{m},\end{equation}
where $\chi\equiv\frac{\left|V_{int}\right|}{2\pi^{2}}\frac{1}{4T_{c}}\int_{0}^{K}\sin^{2}2\theta_{k}k^{2}dk$
encapsulates the overall effect of the quasi-particle states on the
gap response. ($K$ is the upper limit of the momentum summation required
by the standard renormalization procedure.) In the considered regime,
$\chi\simeq\frac{k_{F}\left|a\right|}{2}\frac{\Delta_{0}}{T_{c}}\ll1$
\cite{key-22}, where $k_{F}$ is the Fermi wave number and $a$ is
the background scattering length. Eq. (9) along with Eq. (3) for the
molecular field constitute a closed set of equations for the system
evolution. From them, it is apparent that $\delta\Delta$ is determined
by the combination of effective driving, coming from the term $\frac{\chi}{\tau_{f}}g\delta\phi_{m}$,
and damping with rate $\frac{1-\chi}{\tau_{f}}$. As the driving is
continuously taking the system out of equilibrium, the relaxation
mechanism is permanently activated. The combined effect of both mechanisms
can be expected to produce a nondirect following to the external field.
Approximate analytical solutions confirm these predictions: the gap
evolution is given by \begin{eqnarray}
\frac{\Delta(t)}{\Delta_{0}} & =1 & -C\left[e^{-t/\tau_{R}}\sin\varphi+\sin(\omega_{p}t-\varphi)\right],\end{eqnarray}
where $C=2\chi\frac{g^{2}}{\left|V_{int}\right|}\frac{J_{1}(A/\omega_{p})}{\omega_{p}\sqrt{1+(\omega_{p}\tau_{f})^{2}}}$($>0$)
determines the amplitude of the induced oscillations, $\varphi=\arctan(\omega_{p}\tau_{f})$
is a phase shift with respect to the applied magnetic field, and $\tau_{R}=\frac{\tau_{f}}{1-\chi}$
appears as the condensate relaxation time. (The meaning of $\tau_{R}$
becomes evident in a simplified scenario: for the system with no external
driving, a sudden perturbation of the gap is shown to relax to equilibrium
with characteristic time $\tau_{R}$.) In the considered regime, namely,
near the critical temperature and for a perturbative gap variation,
it is found that $\tau_{R}\simeq\tau_{f}$. The correspondence of
these results with the experimental findings is summarized in the
following points. 

(i) The system response contains a transitory decay with characteristic
time $\tau_{R}$, and, as observed in the experiments, a secular oscillatory
behavior with $\omega_{p}$. The amplitude, which combines in a nontrivial
way parameters of the external field and characteristics of the unperturbed
system, reflects the complex character of the driving mechanism. The
following to the external field is not instantaneous: there is a delay
time associated with the phase shift $\varphi$ and given by $\tau_{D}=\frac{\varphi}{\omega_{p}}=\tau_{R}\left[1+\mathcal{O}\left((\omega_{p}\tau_{R})^{2}\right)\right]$.
Hence, as conjectured in Ref. {[}13{]}, $\tau_{D}$ approximately
corresponds to the condensate relaxation time. The delay presents
no changes at different cycles of the external field. Furthermore,
the small magnitude of the correction $\mathcal{O}\left((\omega_{p}\tau_{R})^{2}\right)$
for the conditions of the experiments explains the detected invariance
of the delay scale with the modulation frequency. As reflected by
the minus sign before $C$, there is an extra phase shift $\pi$ between
the gap oscillation and the driving field. This corresponds exactly
to the results presented in Figs. 2 and 4 in Ref. {[}13{]}, where
an inversion of the magnetic-field axis was introduced to facilitate
the observation of the delay. 

(ii) The connection between the different time scales is uncovered.
Since $\tau_{R}$ gives the time for the condensate to reach the thermal
equilibrium, it is directly related to the thermalization rate of
the populations. Notice that $\tau_{R}$ can significantly differ
from $\tau_{f}$ outside the considered regime with $\chi\ll1$. This
can be understood taking into account the intricate interdependence
of the gap and the populations, which implies, in general, a complex
nonlinear contribution of the populations to the gap relaxation \cite{key-14}. 

(iii) The mechanism responsible for the deferred response is rooted
in the finite reaction time of the gas to a variation in the quasiparticle
energies. {[}Note that the adiabatic limit corresponding to a sudden
relaxation, i.e., to $\tau_{f}\rightarrow0$, is consistently recovered
in Eq. (10).{]} Moreover, as conjectured in a preliminary analysis,
the observed delay is a temperature effect.  At zero temperature,
there is no initial population of the excited states; furthermore,
as the modulation in the assumed regime does not induce a transfer
from the fundamental state, the excited states are never populated.
Hence, at $T=0$, there is no population relaxation, and, consequently,
no delay in the gap evolution \cite{key-23}.

(iv) It is worth discussing the effects that can be expected outside
the considered regime of nonexciting frequencies ($\omega_{p}<2\tilde{\Delta}_{0}$)
and perturbative amplitudes. First, for $\omega_{p}\geq2\tilde{\Delta}_{0}$,
the magnetic field can induce an effective resonance between the vacuum
state and a doubly-excited state. As a consequence, the interaction
between those states, represented by the non-diagonal terms in Eq.
(4), becomes important, and significant heating can result. (See Refs.
{[}24{]} and {[}25{]} for related work.) Second, as the amplitude
is increased, the contribution of the terms of order higher than one
in the expansion of the exponential in Eq. (2) grows. Given that the
frequency of each term is a multiple of $\omega_{p}$, a resonance
between the ground state and the excitations can eventually be reached
for increasing order, which, again, can lead to an irreversible loss
of population.

At this point some aspects of our approach must be recalled. Importantly,
a perturbative regime has been considered: the system, which is initially
in the BCS side, is assumed to be inside that regime during the whole
modulation process. Hence, the unitary limit of large scattering length
is never reached and neither is attained the BEC side. Our approach,
directly set up from the fundamental theory, has some similarities
with former studies in superconductivity where  the effect of a population
disequilibrium on the gap dynamics was tackled by introducing an operative
changing temperature in the static Ginzburg-Landau (GL) equation \cite{key-14,key-15}.
The effective time-dependent GL equation thus obtained was shown to
satisfactorily explain the relaxation process. That description, as
ours, is basically built from the incorporation of formal solutions
for the evolving populations into the gap equation. In the uniform
case and in the perturbative regime for the gap variation, the analogy
with our self-contained approach is complete. Incidentally, we stress
that the success of our uniform description in reproducing the experimental
results, which, in fact, were obtained for a harmonic confinement,
reflects the robustness of the identified physical mechanisms against
spatial non-uniformities. The generalization through a local-density
approximation is straightforward.

In summary, we have presented an analytical explanation for the delayed
response of a Fermi condensate to a modulation of the interaction
strength. Although a more quantitative comparison with the experiments
of Ref. {[}13{]} requires additional information on the temperature
and the amplitude of the applied field, the study uncovers fundamental
aspects of the out-of-equilibrium dynamics of the condensate, in particular,
the nontrivial role of the relaxation in a time regime of experimental
and theoretical interest. Our analysis can have direct practical implications.
One of the main motivations for the experiments was the validation
of the detection schemes based on a projection of the Fermion pair
condensate into a molecular condensate. Those schemes are applicable
only if the response time of the condensate to variations in the interaction
strength is much larger than the sweep time. The relevance of the
nonzero-temperature character of the delay to the projection techniques
is clear: an out-of-equilibrium situation and the subsequent relaxation
process of the condensate can be induced not only by changing the
temperature but also by manipulating the system with external fields.
Our picture provides the theoretical basis for the design of methods
for measuring the different time scales, and, consequently, for defining
the appropriate ranges for the projection. Furthermore, the identification
of the dominant mechanisms opens the way to develop variations of
the basic arrangement as probing tools for different aspects of the
dynamics. Given the generality of the applied model, the applicability
of the study in parallel contexts can be expected.

\end{document}